\documentclass[aps,twocolumn,showpacs,amsmath,floatfix]{revtex4}
\usepackage{graphicx}
\begin{document}

\title{Three-dimensional solitons and vortices in dipolar Bose-Einstein condensates}
\author{Yu.A. Zaliznyak$^1$, A.I. Yakimenko$^{1,2}$}
\affiliation{$^1$Institute for Nuclear Research, Kiev 03680, Ukraine \\
$^2$Department of Physics, Taras Shevchenko National University,
Kiev 03022, Ukraine}
\begin{abstract}
Three-dimensional solitary and vortex structures in Bose-Einstein
condensates are studied in the framework of Gross-Pitaevskii model
including the simultaneous action of local cubic-quintic
nonlinearity and nonlocal dipole-dipole interactions. Nonlocal
interactions are shown to change significantly the formation
threshold and the numbers of atoms confined into the coherent
structures. An appearance of robust high-order ($m=2$)
three-dimensional vortices is revealed.
\end{abstract}
\maketitle

\section{Introduction}
Recent progress in experimental and theoretical studies of
Bose-Einstein condensates (BECs) has opened up a new opportunity
to investigate nonlinear interactions of atomic matter waves
\cite{Petchik,KivsharAgrawal}. By applying an external magnetic
field and using the Feshbach resonance, the $s$-wave scattering
length can be tuned, thus it is possible to explore extreme
regimes of interaction from strongly repulsive to strongly
attractive. The resulting nonlinear evolution of matter waves
gives the possibility to observe the nonlinear effects such as
atomic self-focusing and formation of solitons.

Dark \cite{Burger} and bright solitons
\cite{Strecker,Khaykovich,Kevrekidis} have already been created in
BECs. In the absence of an external trap, such 3D structures are
always unstable: they either collapse, if number of atoms exceeds
some critical value, or spread out in the opposite case. During an
implosion of BEC, the atom density becomes high, and repulsive
three-particle interactions come into play and should be taken
into account, which gives rise to the additional local quintic
nonlinear term  in corresponding Gross-Pitaevskii equation (GPE)
\cite{KovalevKosevich1976,KosevichIvanov90}. In Ref.
\cite{MihalachePRL02} 3D non-spinning solitons and vortex solitons
have been studied in the framework of nonlinear cubic-quintic (CQ)
Scr\"odinger equation with the application to the bulk nonlinear
optical media. The competition between the cubic attractive and
quintic repulsive terms is able to arrest the collapse and to
stabilize 3D solitons. In contrast to non-spinning solitons,
vortex solitons can be unstable with respect to azimuthal
perturbations which brake 3D vortex into several filaments. In
conservative CQ media, single-charged 3D vortices can be
stabilized, while higher-order vortices remain unstable
\cite{MihalachePRL02}. Very recently, stable double-charge vortex
solitons were found in the framework of the complex
Ginzburg-Landau equation \cite{MihalachePRL06}. As the matter of
fact, dissipative solitary structures may be more compact and more
robust, thus it may be easier to find stable localized vortices in
dissipative systems than in their conservative counterparts. In
the present paper we present the first example of stable 3D
spinning higher-order ($m>1$) solitons in a conservative nonlinear
system. The stabilization is achieved due to the action of the
{\it nonlocal} nonlinearity associated with long-range
interparticle interactions in dipolar BEC.


Nonlocal nonlinear media response naturally appears in a wide
variety of physical systems such as plasmas
\cite{LitvakSJPlPhys75,Davydova95,MyPRE05}, Bose- Einstein
condensates BEC \cite{PedriSantosPRL}, optical media
\cite{KrolikovskiJOptB04}, liquid crystals
\cite{RassmusenPRE05,ContiPRL03}, and soft matter
\cite{ContiPRL05}. In the 2D case, nonlocal nonlinear interactions
were shown to suppress an azimuthal instability and to stabilize
vortex solitons \cite{Krolikowski,MyPRE06, MyPRE05, LashkinBEC}.
It was found recently \cite{MihalacheThermal3dPRE06} that spinning
3D solitons are unstable against splitting into a set of stable
fundamental solitons in the medium with nonlocal thermal
nonlinearity.

In degenerate dipolar gases, the nonlocal nonlinearity arises from
long-range, partially attractive, and anisotropic dipole-dipole
(DD) interactions. The theoretical investigations have shown that
the stability of dipolar gases crucially depends on the trap
geometry \cite{Yi00,Santos}. However, it is not always necessary
to include an external trap since self-action of BEC can be enough
to provide the conditions for formation of stable localized
coherent structures.
 In the present paper we address the
following question: how the nonlocal DD interactions affect the
parameters and stability properties of the self-consisted solitons
and vortex solitons formed in the trap-free BEC.

\section{Model equations}
We consider BEC of atoms with electric dipole $d$ (the model is
equally valid for magnetic dipoles) oriented in the $z$ direction
by a sufficiently strong external field, and that hence interact
via a dipole-dipole potential
\begin{equation}\label{eq:Vd}
V_d(\textbf{r})=g_d(1-3\cos^2\vartheta)/r^3,\end{equation} where
$g_d=\alpha d^2/4\pi\varepsilon_0$, $\varepsilon_0$ is the vacuum
permittivity, $\vartheta$ is the angle formed by the vector
joining the interacting particles and the dipole direction. The
parameter $\alpha$ can be tuned (see e.g. \cite{PedriSantosPRL})
in the range $-1/2\le\alpha\le 1$.

In the mean field approximation, a dipolar BEC at sufficiently low
temperatures is described by a GPE with nonlocal nonlinearity:
\begin{eqnarray}
 \label{GPEqNonStationary}
 \nonumber
   i\hbar\frac{\partial\Psi}{\partial t} + \frac{\hbar^2}{2M}\Delta\Psi
   -g \Psi \left|\Psi\right|^2
   +g_K \Psi \left|\Psi\right|^4 \\ + g_d \Psi\int  V_d(\textbf{r} -\textbf{r}')
   |\Psi(\textbf{r}', t)|^2 d^3\textbf{r}\,' = 0,
\end{eqnarray}
where $g =4\pi\hbar^2a/M$, $a$ is the $s$-wave scattering length.
In the following we consider attractive two-particle interaction
($a<0$) and repulsive three-particle interaction ($g_K<0$). Note
that we account here only for the conservative part of
three-particle interaction. Thus, GPE (\ref{GPEqNonStationary})
conserves the norm (the number of atoms in the condensed state):
\begin{equation}
 \label{N3D}
 N = \int\left| \Psi \right|^2 d^3 \textbf{r},
\end{equation}
energy:
\begin{eqnarray}  \label{Hamilt}
\nonumber E = \int\left\{\frac{\hbar^2}{2M}| \nabla \Psi|^2
+\frac{1}{2}g|\Psi|^4 -\frac{1}{3}g_K|\Psi|^6\right.
\\ \left.-\frac12g_d|\Psi|^2\Theta \right\}
d^3\textbf{r},\, \Theta=\int  V_d(\textbf{r}
-\textbf{r}')|\Psi(\textbf{r}')|^2d^3\textbf{r}',
\end{eqnarray}
momentum and angular momentum.

In the next section, the general properties of stationary solitons
and vortices are studied by analytical variational method and
numerically.

\section{Stationary 3D solitons and vortex solitons}
Stationary solutions of the Eq. (\ref{GPEqNonStationary}) have the
form $\Psi(\textbf{r},t)=\psi(\textbf{r})\exp(i\lambda t)$ and
obey the dimensionless equation:
\begin{eqnarray}
 \label{GPEqStationary}
   -\lambda \psi + \Delta\psi + \psi \left|\psi\right|^2
   - \psi \left|\psi\right|^4 + C \psi\,  \Theta = 0,
\end{eqnarray}
\begin{equation}\label{eq:Theta}
\Theta =\hat F^{-1}\left[\tilde{V}_d(k)\hat
F\left\{|\psi|^2\right\}\right],
\end{equation}
where $\hat F$ denotes the Fourier transformation,
$\tilde{V}_d(k)=\frac{4\pi}{3}(3k_z^2/k^2- 1)$ is the Fourier
transform of dipole-dipole interaction potential. Applying the
following rescaling transformations: $\textbf{r}_{\rm dimless}=
(2Mg^2/\hbar^2g_K)^{1/2}\textbf{r}_{\rm phys}$, $\psi_{\rm
dimless}= |g_K/g|^{1/2}\psi_{\rm phys}$, the number of free
parameters is reduced to two, namely $C=g_d/g$, and
$\lambda=-\mu|g_K|/g^2$ ($\mu$ being the chemical potential).

In order to study the general properties of stationary 3D solitary
and vortex solutions analytically, we employed the variational
analysis with the following ansatz:
\begin{equation}\label{eq:ansatz}
\Psi(\textbf{r},t)=h\left(\frac{\rho}{a_\rho}\right)
^{|m|}\exp\left\{{-\frac{\rho^2}{2a_\rho^2}-\frac{z^2}{2a_z^2}+i
m\varphi}\right\},
\end{equation}
where $\rho=\sqrt{x^2+y^2}$, $\varphi$ is azimuthal angle. Integer
$m$ is the topological charge, $m=0$ corresponds to non-spinning
solitons, $m>0$ -- to vortex solitons. Two variational parameters
$a_\rho(t)$ and $a_z(t)$ describe radii of the structure in the
directions across and along the external field, and the amplitude
$h(t)=N^{1/2}\pi^{-3/4}a_\rho^{-1}(t)\left[a_z(t)
m!\right]^{-1/2}$ is found from normalization condition
(\ref{N3D}). Using the trial function (\ref{eq:ansatz}) one can
obtain the energy functional (\ref{Hamilt}) at fixed number of
atoms $N$ as the function of two variational parameters $a_\rho$
and $a_z$. These calculations were performed analytically, but the
resulting expressions are too cumbersome to be presented here.
Stationary solutions correspond to the stationary points of the
energy functional $E$, i.e. the parameters $a_\rho$ and $a_z$ are
found from the set of equations: $\partial E/\partial a_z$=0,
$\partial E/\partial a_\rho=0$ at fixed $N$. The dependencies
$N(\lambda)$ obtained by variational method below are compared
with the numerical results.

\begin{figure}[h]
\centerline{\includegraphics[width=3.4in]{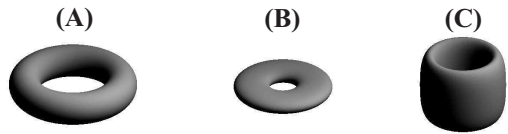}}
\caption{Numerically found stationary vortex solutions
(isosurfaces of constant $|\psi|^2$) for $m=2$ and different
values of the nonlocality parameter $C$: (A) $C=0$, $\lambda =
0.1$, (B) $C=0.3$, $\lambda = 0.45$, (C) $C=-0.3$, $\lambda =
0.08$.} \label{fig:3DVortices}
\end{figure}

To find the stationary solutions of Eq. (\ref{GPEqStationary})
numerically, the stabilized iteration procedure similar to that
proposed by Petviashvili (see, e.g. \cite{Petviashvili86}) was
implemented on a fully 3D grid with the resolutions up to $128^3$.
Figure \ref{fig:3DVortices} shows the isosurfaces of constant
$|\psi|^2$ for vortex solitons $m=2$ at different $C$. The change
in sign of nonlocality parameter $C$ leads to flattening ($C>0$)
or to elongation ($C<0$) of the distribution of atoms along
$z$-direction compared to the case $C=0$. Below we shall
concentrate mainly at the case $C>0$.

\begin{figure}[h]
\centerline{\includegraphics[width=3.4in]{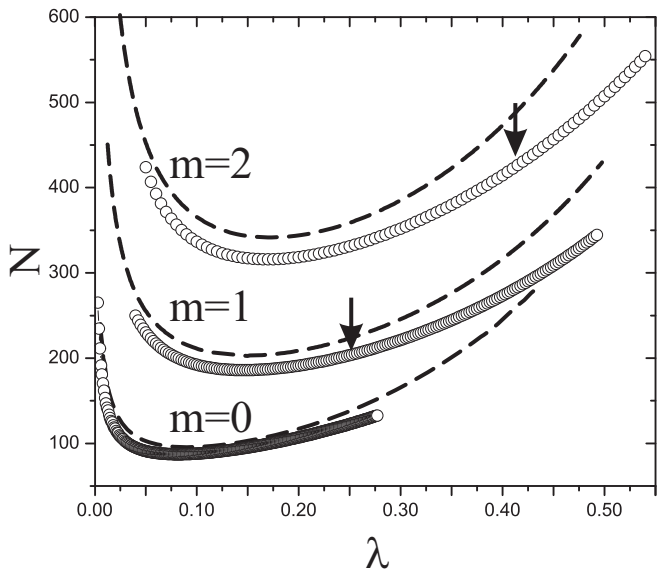}}
\caption{Number of particles $N$ vs $\lambda$ for different
stationary solutions at $C=0.3$. Numerical results are shown in
circles, dashed lines are for variational predictions. The arrows
indicate stability thresholds for vortices.} \label{fig:EDD1}
\end{figure}

\begin{figure}[h]
\centerline{\includegraphics[width=3.4in]{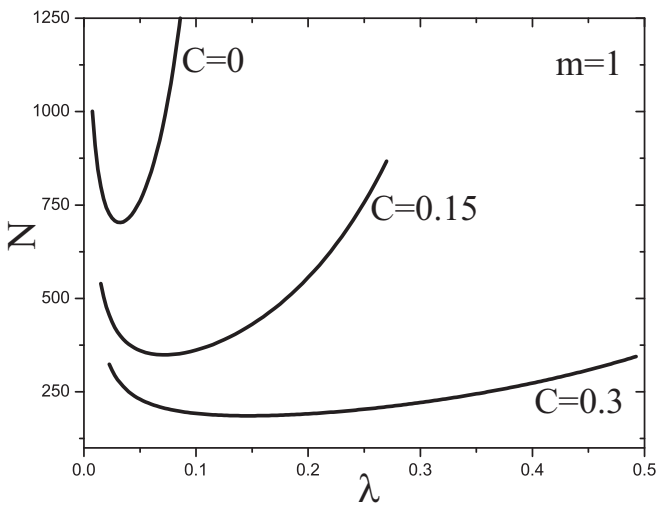}}
\caption{Number of particles $N$ vs parameter $\lambda$ for
vortices with $m=1$ with different $C$. Numerical results.}
\label{fig:EDD2}
\end{figure}

In the Fig.  \ref{fig:EDD1} the numbers of atoms versus the
parameter $\lambda$ for solitons ($m=0$) and vortices ($m=1,2$)
are shown. The numerically found states are plotted by circles,
while the dashed lines present analytical dependencies calculated
by the variational method. It is seen that the variational
predictions are in a good agreement with the results of numerical
simulations. The divergence between analytical and numerical
curves increases as the number of atoms $N$ confined into the
structure  goes up, and this can be explained by the observation,
that with increase of $\lambda$, soliton and vortex solutions
develop more and more pronounced plateau and their profiles then
strongly deviate from the trial Gaussian profile used in the
variational approach. In the Fig. \ref{fig:EDD2}, numbers of atoms
are plotted versus the parameter $\lambda$ for different strengths
of non-locality for the vortices $m=1$ (dependencies $N(\lambda)$
for solitons $m=0$ and vortices $m=2$ show very much similar
behavior). One can see that increased nonlocality leads to the
significant decrease of the structure's formation threshold and to
the substantial elongation of the range of accessible parameter
$\lambda$.



\section{Stability and dynamics}

We start investigation of the stability of vortices with analysis
of small perturbations applied to the stationary solution:
$$\Psi=(\psi +\epsilon)e^{i\lambda t},$$
where $|\epsilon(\textbf{r},t)|<<|\psi|$. Linearizing the GPE
(\ref{GPEqNonStationary}) in vicinity of stationary solution one
gets the nonstationary equation which describes evolution of small
perturbation:
\begin{eqnarray}
\nonumber i\frac{\partial\epsilon}{\partial
t}-\lambda\epsilon+\Delta\epsilon+(2|\psi|^2\epsilon
+\psi^2\epsilon^*)\\ \label{eq:LinEqEpsilon}
-(2\psi^2\epsilon^*+3|\psi|^2\epsilon)|\psi|^2
+C(\delta\Theta\,\psi+\Theta\epsilon)=0,
\end{eqnarray}
where $\Theta$ is given by Eq. (\ref{eq:Theta}) and
$$
\delta\Theta=\hat F^{-1}\left[\tilde{V}_d(k)\hat
F\left\{\psi\epsilon^*+\psi^*\epsilon\right\}\right].$$ We have
solved Eq. (\ref{eq:LinEqEpsilon}) numerically by means of
split-step technic. The azimuthal instability growth rate was
calculated as follows:
$$\gamma=\frac{1}{2\Delta t}\ln\left\{\frac{\nu(t+\Delta t)}{\nu(t)}\right\},$$
where $\Delta t$ is the time step, and
$\nu(t)=\langle\epsilon|\epsilon\rangle$ is the norm of the
perturbation. When the perturbation grows exponentially,
$\gamma(t)$ saturates at some value $\gamma$, which was taken as
the maximum growth rate.

\begin{figure}[h]
\centerline{\includegraphics[width=3.4in]{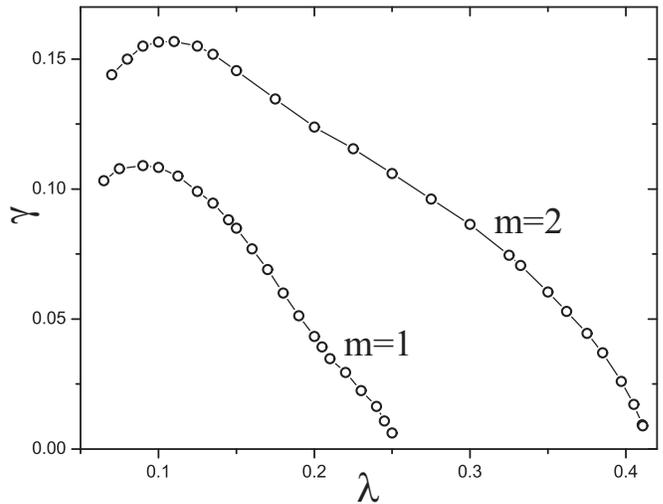}}
\caption{Maximum growth rates $\gamma$ vs $\lambda$ for vortices
with $m=1$ and $m=2$, C=0.3} \label{fig:GrRates}
\end{figure}

The soliton structures having zero topological charge are
azimuthally stable, and their stability region coincides with that
described by the Vakhitov-Kolokolov criterion ($\partial
N/\partial\lambda>0$) \cite{VakhitovKolokolov}. The maximum growth
rates obtained by means of above approach are given in Fig.
\ref{fig:GrRates} for vortices $m=1$ and $m=2$ at $C=0.3$.

Note that our analysis is formulated in very general form, it
gives the {\it maximum} growth rate at given $\lambda$ but not the
growth rate of specific unstable azimuthal mode. This explains the
knees in the dependencies $\gamma(\lambda)$ (see Fig.
\ref{fig:GrRates}) which correspond to the intersections of growth
rates $\gamma_L(\lambda)$ for different azimuthal numbers $L$ of
perturbations.

As it is known (see, e.g., \cite{MyPRE2003}),  the competition
between cubic and quintic nonlinear terms leads to the formation
of solitary structures having a kind of plateau on the top when
number of atoms $N$ is high enough. Further increase of $N$ leads
to saturation of the amplitude and to elongation of the plateau.
In the previous works \cite{MyJoPA,MichinelPRL06}, stabilization
of the 2D vortices was associated with the appearance of such
plateau. It is remarkable that in the BECs described by the GP
model with additional nonlocal nonlinearity, stabilization of
vortices occurs {\it before} the plateau is formed. At given
nonlocality parameter $C$, the growth rates vanish at some
threshold value of $\lambda$ due to the action of nonlocal
dipole-dipole interactions. For instance, at $C=0.3$, the
single-charged vortices get the stability window at $\lambda>
0.26$ and double-charged vortices ($m=2$) -- at $\lambda > 0.42$,
these threshold values are marked by the arrows in Fig.
\ref{fig:EDD1}. As for vortices with $m>2$, no stabilization was
observed.

The evolution and the dynamical stability of solitary and vortex
structures was simulated numerically using the well-known
split-step Fourier method \cite{AgrawalBook}, where the nonlocal
DD integral term was calculated in the spectral space. We used
perturbed solutions found by the stationary solver as an initial
condition for the nonstationary problem.  For the structures with
plateau, Petviashvili's approach may fail and is unable to
reproduce the $N(\lambda)$ diagram completely in the high-energy
region. In this case, we used the appropriate initial condition
with the parameters extracted from the variational analysis as an
input for the non-stationary solver to check an existence and
stability of solitons and vortices for values of $N$ where our
relaxative solver is inefficient.
\begin{figure}[htb]
\centerline{\includegraphics[width=3.4in]{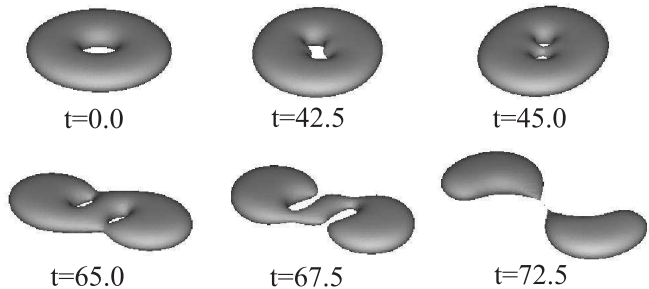}}
\caption{Decay of unstable  vortex ($m=2$, $C=0.3$,
$\lambda=0.25$): isosurfaces of constant particle density
$|\psi|^2$ are given for different times.} \label{fig:Evolution1}
\end{figure}
\begin{figure}[htb]
\centerline{\includegraphics[width=3.4in]{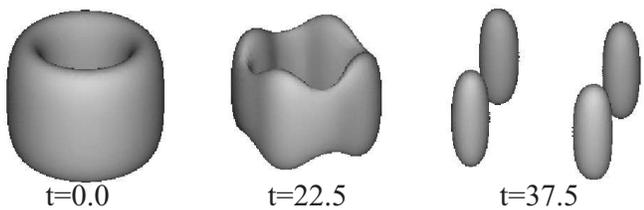}}
\caption{Decay of unstable  vortex ($m=2$, $C=-0.3$,
$\lambda=0.08$): isosurfaces of constant particle density
$|\psi|^2$ are given for different times.} \label{fig:Evolution2}
\end{figure}

Direct numerical simulations confirmed the predictions obtained by
the linear stability analysis. Solitons were found to be stable
everywhere in the region $\partial N/ \partial \lambda > 0$,
stability thresholds for vortices coincide with those predicted by
the linear stability analysis. In the unstable region, destruction
of a vortex may occur in the different ways, depending on the
number of atoms confined into the structure and on the applied
perturbation. Typical unstable evolution snapshots are plotted in
Fig. \ref{fig:Evolution1} and Fig. \ref{fig:Evolution2} for
different signs of nonlocality parameter $C$.

\section{Conclusions}

In the present paper we study the influence of nonlocal
dipole-dipole interparticle interaction on 3D solitons and
vortices in BEC described by GPE with local contact attractive
two-particle and repulsive three-particle interaction. Nonlocal
anisotropic dipole-dipole interaction leads to quantitative, as
well as to qualitative changes in scales, energies and stability
properties of supported solitary structures. The formation
threshold (minimum number of atoms, which is needed to form a
structure) goes down compared to the case of absence of DD
interactions, and possible range of chemical potentials
significantly elongates. While soliton structures having zero
topological charge are stable if $\partial N/\partial\lambda>0$,
vortices may exhibit azimuthal instability. We proved the
stabilization of 3D vortices with $m=1$ and $m=2$ and revealed
their stabilization thresholds. All higher order ($m>2$) vortices
were found to be unstable. Up to our knowledge, this is a first
example of 3D stable vortex with the topological charge larger
than one in a conservative system.

\begin{acknowledgments}
We are grateful to Yuri Kivshar and Volodymyr Lashkin for fruitful
discussions and comments about this paper. \end{acknowledgments}

\end{document}